\documentclass[aps,showkeys,twocolumn,preprintnumbers,amsmath,amssymb,superscriptaddress,floatfix,nofootinbib]{revtex4}
\usepackage{graphicx,color,dcolumn,booktabs,bm}
\usepackage{longtable,lscape}
\usepackage{txfonts}
\usepackage{overpic}
\usepackage{epsfig}
\usepackage{amssymb}
\usepackage{rotating}
\usepackage{epstopdf}
\usepackage{appendix}
\usepackage{indentfirst}
\usepackage{feynmf}   
\usepackage{slashed}  
\usepackage{cases}
\usepackage{color}
\usepackage{multirow}
\usepackage{graphicx,color,dcolumn,booktabs,bm}
\usepackage{cases}
\usepackage{array}
\usepackage{float}
\usepackage{subfigure}

\graphicspath{{Figures/}} %

\usepackage[colorlinks, citecolor=blue,anchorcolor=red,menucolor=red, linkcolor=red,filecolor=red,runcolor=red,urlcolor=blue,frenchlinks=red]{hyperref}
\usepackage{cleveref}

\begin{document}
\title{Light meson emissions of the selected charmonium-like states within compact tetraquark configurations }
\author{Ning Li}
\affiliation{ Department of Physics, Hunan Normal University, and Key Laboratory of Low-Dimensional Quantum Structures and Quantum Control of Ministry of Education, Changsha 410081, China }

\affiliation{Key Laboratory for Matter Microstructure and Function of Hunan Province, Hunan Normal University, Changsha 410081, China}

\author{Hui-Zhen He}
\affiliation{ Department of Physics, Hunan Normal University, and Key Laboratory of Low-Dimensional Quantum Structures and Quantum Control of Ministry of Education, Changsha 410081, China }

\affiliation{Key Laboratory for Matter Microstructure and Function of Hunan Province, Hunan Normal University, Changsha 410081, China}

\author{Wei Liang}
\affiliation{  School of Physics and Electronics, Hunan Key Laboratory of Nanophotonics and Devices, Central South University, Changsha 410083, China }

\author{Qi-Fang L\"u \footnote{Corresponding author} } \email{lvqifang@hunnu.edu.cn} %
\affiliation{ Department of Physics, Hunan Normal University, and Key Laboratory of Low-Dimensional Quantum Structures and Quantum Control of Ministry of Education, Changsha 410081, China }

\affiliation{Key Laboratory for Matter Microstructure and Function of Hunan Province, Hunan Normal University, Changsha 410081, China}

\affiliation{Research Center for Nuclear Physics (RCNP), Ibaraki, Osaka 567-0047, Japan}

\author{Dian-Yong Chen } \email{chendy@seu.edu.cn} %
\affiliation{ School of Physics, Southeast University, Nanjing 210094, China }

\affiliation{Lanzhou Center for Theoretical Physics, Lanzhou University, Lanzhou 730000, P. R. China }

\author{ Yu-Bing Dong } \email{dongyb@ihep.ac.cn} %
\affiliation{ Institute of High Energy Physics, Chinese Academy of Sciences, Beijing 100049, China }

\affiliation{ Theoretical Physics Center for Science Facilities (TPCSF), CAS, Beijing 100049, China }

\affiliation{School of Physical Sciences, University of Chinese Academy of Sciences, Beijing 101408, China }

\begin{abstract}
We adopt the quark pair creation model to investigate  the light meson emissions of  several  charmonium-like states.   The quark pair creation model is applied to the four-body systems, and  we calculate the pion/kaon emissions of the $X(4700)$, $Z_c(4430)$, $Y(4230)$, $Y(4360)$, $Y(4390)$ and $Y(4660)$ within the compact tetraquark assumptions. It is found that the pion/kaon decay widths for the $X(4700)$ and  $Z_c(4430)$ are relatively  small, while the partial decay widths for the resonances $Y(4230)$, $Y(4360)$, $Y(4390)$ and $Y(4660)$ are significant. We expect that our exploration of these decay behaviors can provide useful  information for future experimental searches and theoretical interpretations.

\end{abstract}


\maketitle

\section{Introduction}
In the past two decades, a large number of new hadronic states have been observed by the large-scale scientific  facilities, and some of them cannot be categorized into the conventional mesons or baryons. These states that do not meet the expectations for conventional hadrons in the quark models are called as exotic states. The theorists have done plenty of works and  attempted to explain the mysterious properties of these exotica. The experimental and theoretical efforts makes the study of these  exotic states as an intriguing field in hardonic physics. These achievements and progresses provide good opportunities for us to investigate the internal structures of exotica, and recent experimental and theoretical status can be found in recent reviews~\cite{Hosaka:2016pey,Chen:2016qju,Lebed:2016hpi,Dong:2017gaw,Guo:2017jvc,Olsen:2017bmm,Liu:2019zoy,Brambilla:2019esw,Barabanov:2020jvn,Chen:2022asf}.

Among those new hadronic states, the charmonium-like states are particularly fascinating, which attract wide attentions both experimentally and theoretically. Various approaches including the phenomenological quark models, effective field theories, QCD sum rule, and lattice QCD have been adopted to study their properties, and some different interpretations have been proposed to reveal their natures, such as compact tetraquarks, meson-meson molecular states, conventional charmonium, hybrid states, kinematic effects and so on. Also, these investigations for charmonium-like states cover the masses, decay behaviors, production processes, and magnetic moments. Although the  previous works have not provided a unified explanation, it is essential to explore every aspect of these states with different configurations, which  can help us to better understand them and search for more undetected particles.

Besides the mass spectra and productions, strong decay behaviors also reflect the internal structures of charmonium-like states, and have been drawing wide attentions increasingly. In the literature, the two-meson decay channels are widely studied~\cite{Yue:2022gym,Sundu:2018toi,Chen:2015fsa,Wang:2019iaa,Chen:2019wjd,Ke:2013gia,Li:2013yla,Liu:2014eka,Wang:2018pwi,Xiao:2019spy,Ferretti:2020civ,Liu:2016nbm,Wang:2022dfd}. For the molecular states, these decays  usually occur through the meson exchange processes, and the fall-apart mechanism is generally assumed to estimate the partial decay widths for the compact tetraquarks. Also, a charmonium-like state can decay into one light meson plus one charmonium-like states, and yet the studies on this kind of decays are few. Under the molecular hypothesis, the authors calculated the  processes with the  effective  Lagrangian approach~\cite{Chen:2017abq,Chen:2016byt}. However, these processes have not been investigated within compact tetraquark configurations.
In fact, the light meson emissions of these $XYZ$ states not only help us to distinguish between the compact tetraquarks and loosely bound  molecules, but also  play an essential role as the connecting bridges among different  charmonium-like states. Hence, it is time to study these decay processes for charmonium-like states within compact tetraquark configurations seriously.  

Experimentally, there have been several  observed candidates for the excited compact tetraquarks. In 2016, a $J^{PC }$ = $0^{++} $ structure $X(4700)$ was  discovered in $ B^{+} \rightarrow J/\psi \phi K^{+}$ decay process by  the LHCb Collaboration ~\cite{LHCb:2016axx,LHCb:2016nsl}, which has a  mass of $4704 \pm 10^{+14}_{-24}$ MeV and a  width of $120 \pm 31^{+42}_{-33}$ MeV. Since the $X(4700)$ was found in the  $J/\psi \phi$ invariant mass, it may include  both  $c \bar c$ pair and  $s \bar s$ pair and could be regarded as a radially excited  $ c \bar c  s \bar s $ compact tetraquark~\cite{Chen:2010ze,Lu:2016cwr,Liu:2021xje,Deng:2017xlb}. Another radially excited  candidate is the resonance $Z_{c}(4430)$, which was firstly observed in the $B \rightarrow K \pi^- \phi(2S)$ decays process by  the Belle Collaboration in 2007~\cite{Belle:2007hrb}. Then this resonance was confirmed in the same process by the LHCb Collaboration and  in the  $B^{0} \rightarrow K^{-}\pi^{+}J/\psi$  decay by the  Belle Collaboration~\cite{Belle:2014nuw,LHCb:2014zfx}. Its mass and decay width were remeasured  to be $4478^{+15}_{-18}$ MeV and $181 \pm31$ MeV, respectively~\cite{ParticleDataGroup:2022pth}. For the  $P-$wave candidates, there have been  several $ 1^{--}$  charmonium-like states. For instance, the Y(4260) was reported in the process $e^{+}e^{-} \rightarrow \pi^{+}\pi^{-} J/ \psi $ by the BaBar Collaboration, and confirmed by the Belle and BESIII Collaborations subsequently~\cite{BaBar:2005hhc,Belle:2007dxy,BESIII:2016bnd}.  With the increase of experimental data, the Y(4260) was renamed as Y(4230), which has a mass of $(4219.6 \pm 3.3 \pm 5.1)$ MeV and  a width of $(56.0 \pm 3.6 \pm 6.9)$ MeV, respectively~\cite{ParticleDataGroup:2022pth}. Moreover, the  decay branching fraction $ \Gamma_{e^{+}e^{-}} \mathcal{B}[Y(4230) \rightarrow \pi^{+}\pi^{-}h_c] $ was measured to be $(4.6^{+2.9}_{-1.4} \pm 0.8)$ eV  in the $\pi^{+}\pi^{-}h_c$ channel\cite{BESIII:2016adj}. The Y(4390) was found  in the  $e^{+}e^{-} \rightarrow \pi^{+}\pi^{-} h_c$ cross sections around 4.22 and 4.39 GeV by  BES\uppercase\expandafter{\romannumeral3}  Collaboration \cite{BESIII:2016adj}, with a mass of $4391.5^{+6.3}_{-6.8} \pm 1.0$ MeV and a width of $139.5^{+16.2}_{-20.6} \pm 0.6$ MeV, respectively. This structure was also seen in the  $e^{+}e^{-} \rightarrow  \eta J/\psi$ and $e^{+}e^{-} \rightarrow \pi^{+}\pi^{-}D^+ D^-$   processes~\cite{BESIII:2020bgb,BESIII:2022quc}. The  experimental information  for other $P-$wave candidates $Y(4360)$ and $ Y(4660)$ can be found in the literatures~\cite{BESIII:2016bnd,Belle:2015hcs,Belle:2014wyt,BaBar:2012hpr,Belle:2008xmh,BaBar:2006ait,Belle:2007umv,Belle:2020wtd,Belle:2019qoi}.

In present work, we apply the $^3P_0$ model to the four-body systems and calculate the selected decays of the resonance $X(4700) $, $Z_c(4430)$, $Y(4230)$, $Y(4360)$, $Y(4390)$ and $Y(4660)$. The final states are light mesons plus $Z_c(3985)$, $Z_c(4020)$, $Z_{cs}(3900)$, or $Z_{cs}(4000)$. According to  previously theoretical and experimental works, these  charmonium-like states are good candidates of the compact tetraquarks. Here, we adopt the compact tetraquark configurations to investigate the decay processes among these $XYZ$ states. We found that the pion/kaon decay widths for the $X(4700) $ and $Z_c(4430)$ are relatively small, while the partial decay widths for the $Y(4230)$, $Y(4360)$, $Y(4390)$,and $Y(4660)$ are significant. 

This paper is organized as follows. The formalism of strong decays  for compact tetraquarks in the $^3P_0$ model are introduced in Sec~\ref{model}. We present the numerical results and discussions for the selected charmonium-like states in Sec~\ref{low-lying}. A short summary are given in the last section.

\section{Model}{\label{model}}
In this work, we adopt the $^3P_0$ model to calculate the Okubo-Zweig-Iizuka-allowed strong decays of  compact tetraquarks states. In this model, a quark-antiquark pair with the quantum number $J^{PC}$ =$0^{++}$ is created from the vacuum and then goes into the final states~\cite{Micu:1968mk}. This model has been widely employed to study the strong decays for different kinds of hadronic systems with notable successes~\cite{LeYaouanc:1988fx,LeYaouanc:1977gm,Roberts:1992esl,Ackleh:1996yt,Barnes:1996ff,Barnes:2002mu,Zhao:2016qmh,Chen:2007xf,Chen:2016iyi,Lu:2016bbk,Ferretti:2014xqa,Godfrey:2015dva,Segovia:2012cd,Lu:2020ivo,Liang:2020hbo,He:2021xrh}. Also, some previous works performed the formalism for compact tetraquarks~\cite{Roberts:1992esl,Ader:1979bb,Roberts:1990ky,Liu:2016sip}. In the nonrelativistic limit, to describe the decay process $A\rightarrow BC$ for a compact tetraquark, the transition operator $T$  in the $^3P_0$ model can be expressed as \cite{Liang:2020hbo,He:2021xrh,Zhao:2016qmh,Chen:2007xf,Lu:2020ivo,Lu:2016bbk}
\begin{eqnarray}
T&=&-3\gamma\sum_m\langle 1m1-m|00\rangle\int
d^3\boldsymbol{p}_5d^3\boldsymbol{p}_6\delta^3(\boldsymbol{p}_5+\boldsymbol{p}_6)\nonumber\\&&\times {\cal{Y}}^m_1\left(\frac{\boldsymbol{p}_5-\boldsymbol{p}_6}{2}\right)\chi^{56}_{1,-m}\phi^{56}_0\omega^{56}_0b^\dagger_{5i}(\boldsymbol{p}_5)d^\dagger_{6j}(\boldsymbol{p}_6),
\end{eqnarray}
where $\gamma$ is a dimensionless $q_5\bar{q}_6$ quark pair production strength, and $\boldsymbol{p}_5$ and $\boldsymbol{p}_6$ are the momenta of the created quark $q_5$ and antiquark  $\bar{q}_6$, respectively. Here, the $i$ and $j$ are the color indices of the created quark and antiquark, $\phi^{56}_{0}=(u\bar u + d\bar d +s\bar s)/\sqrt{3}$, $\omega^{56}=\delta_{ij}$, and $\chi_{{1,-m}}^{56}$ are the flavor singlet, color singlet, and spin triplet wave functions of the quark-antiquark pair, respectively. The ${\cal{Y}}^m_1(\boldsymbol{p})\equiv|p|Y^m_1(\theta_p, \phi_p)$ is the solid harmonic polynomial reflecting the $P-$wave momentum-space distribution of the $q_5\bar{q}_6$ pair.

\begin{figure}[!htbp]
\includegraphics[scale=0.9]{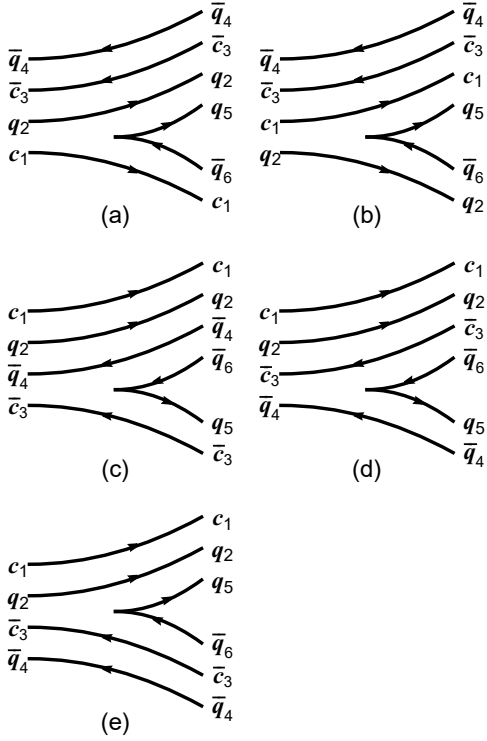}
\vspace{0.0cm} \caption{ The five possible  rearrangements. (a) and (c) are charmed(-strange) meson emissions, (b) and (d) are light meson emissions, and (e) stands for the baryon and anti-baryon decay modes.}
\label{Fig1}
\end{figure}

For the strong decays  of compact tetraquarks $c q \bar c \bar q$, there exist  five possible  rearrangements: 
\begin{eqnarray} 
A(c_1,q_2,{\bar c}_3,{\bar q}_4)+ P(q_5,{\bar q}_6)\rightarrow B(q_2,q_5,{\bar c}_3,{\bar q}_4)+C(c_1,{\bar q}_6)
\end{eqnarray} 
\begin{eqnarray} 
A(c_1,q_2,{\bar c}_3,{\bar q}_4)+ P(q_5,{\bar q}_6)\rightarrow B(c_1,q_5,{\bar c}_3,{\bar q}_4)+C(q_2,{\bar q}_6)
\end{eqnarray}
\begin{eqnarray} 
A(c_1,q_2,{\bar c}_3,{\bar q}_4)+ P(q_5,{\bar q}_6)\rightarrow B(c_1,q_2,{\bar q}_4,{\bar q}_6)+C(q_5,{\bar c}_3)
\end{eqnarray} 
\begin{eqnarray} 
A(c_1,q_2,{\bar c}_3,{\bar q}_4)+ P(q_5,{\bar q}_6)\rightarrow B(c_1,q_2,{\bar c}_3,{\bar q}_6)+C(q_5,{\bar q}_4)
\end{eqnarray} 
\begin{eqnarray} 
A(c_1,q_2,{\bar c}_3,{\bar q}_4)+ P(q_5,{\bar q}_6)\rightarrow B(c_1,q_2,q_5)+C(\bar{c}_3,{\bar q}_4, {\bar q}_6)
\end{eqnarray} 
In Figure~\ref{Fig1}, we present all possible rearrangements including the charmed(-strange) meson emissions, light meson emissions, and baryon-antibaryon processes. Owing to the limited phase space for selected charmonium-like states, we only focus on the light meson emissions in present work.

\begin{figure}[!htbp]
	\centering
	\includegraphics[scale=0.6]{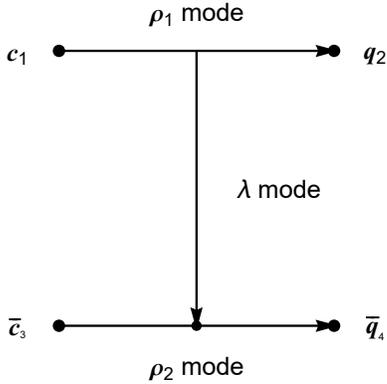}
	\vspace{0.0cm} \caption{The Jacobi coordinates of compact tetraquarks.}
	\label{Fig2}
\end{figure}

 To investigate  the  strong decay behaviors of compact tetraquarks,  we present the Jacobi coordinates  in Figure~\ref{Fig2}. The $c_1$ and $q_2$ stand for quarks, and $\bar c_3$ and $\bar q_4$ correspond antiquarks. The $\boldsymbol \rho_1=\boldsymbol{r_1}-\boldsymbol{r_2}$ is the relative coordinate between two quarks, the $\boldsymbol{\rho_2}=\boldsymbol{r_3}-\boldsymbol{r_4}$ is the relative coordinate between two antiquarks, and the $\boldsymbol \lambda=\frac{m_1\boldsymbol{r_1}+m_2\boldsymbol{r_2}}{m_1+m_2}-\frac{m_3\boldsymbol{r_3}+m_4\boldsymbol{r_4}}{m_3+m_4}$ represents the relative coordinate between quarks and antiquarks. With this definition, we can classify the excitations of compact tetraquarks into three types: $\rho_1-$mode, $\rho_2-$mode and $\lambda-$mode. Empirically, the observations for $\rho-$mode excitations are scare for heavy-light systems. For instance, most observed singly heavy baryons in experiments can be assigned as $\lambda$-mode excitations, and no  $\rho-$mode heavy baryon is confirmed until now. In present work, we only consider the $\lambda$-mode excitations for the initial states, that is $n_{\rho_{A1}}=l_{\rho_{A1}}=n_{\rho_{A2}}=l_{\rho_{A2}}=0$ and $L_A = L_{\lambda_A}$. Also, due to constraint of the  phase space, all the final states are ground  states.

 With this transition operator $T$, the helicity amplitude ${\cal{M}}^{M_{J_A}M_{J_B}M_{J_C}}$ is defined as 
\begin{eqnarray}
\langle
BC|T|A\rangle=\delta^3(\boldsymbol{P}_A-\boldsymbol{P}_B-\boldsymbol{P}_C){\cal{M}}^{M_{J_A}M_{J_B}M_{J_C}}.
\end{eqnarray}
Here, we take the  diagram (d) in Figure~\ref{Fig1} as an example. Similar to the strong decays of conventional hadrons~\cite{Liang:2020hbo,He:2021xrh,Zhao:2016qmh,Chen:2007xf,Lu:2020ivo,Lu:2016bbk}, the explicit formula of  helicity amplitude ${\cal{M}}^{M_{J_A}M_{J_B}M_{J_C}}$  can be expressed as
\begin{eqnarray}
&&{\cal{M}}^{M_{J_A}M_{J_B}M_{J_C}}\nonumber\\
&&=- \gamma \sqrt{8E_AE_BE_C}\sum_{M_{L_A}}\sum_{M_{S_1}}\sum_{M_{S_2}}\sum_{M_{S_3}}\sum_{M_{S_4}}\sum_{m}\nonumber\\
&&\times \langle S_1 M_{S_1} S_2 M_{S_2}|S_{12} M_{S_{12}}\rangle  \langle S_3 M_{S_3} S_4 M_{S_4}|S_{34}M_{S_{34}}\rangle\nonumber \\ 
&&\times \langle S_{12} M_{S_{12}} S_{34} M_{S_{34}}| S_A M_{S_A}\rangle\times \langle L_A M_{ L_A} S_A M_{S_A} |J_A M_{J_A}\rangle\nonumber\\
&& \times \langle S_1 M_{S_1} S_2 |M_{S_2} S_{12} M_{S_{12}} \rangle  \times \langle S_3 M_{S_3} S_6 M_{S_6} |S_{36} M_{S_{36}}\rangle  \nonumber\\
&& \times \langle S_{12} M_{S_{12}} S_{36} M_{S_{36}} |J_B M_{J_B}\rangle \times \langle S_4 M_{S_4} S_5 M_{S_5} |J_C M_{J_C}\rangle \nonumber\\
&&\times \langle S_5 M_{S_5} S_6 M_{S_6} |1 -m\rangle \times \langle 1 m 1 -m |0 0\rangle\nonumber \\
&&\times \langle \phi_B^{1236} \phi_C^{45}|\phi_A^{1234}\phi_0^{56}\rangle I^{M_{L_A}m}_{M_{L_B}M_{L_C}}(\boldsymbol{p}),
\end{eqnarray}
where $\langle \phi_B^{1236} \phi_C^{45}|\phi_A^{1234}\phi_0^{56}\rangle$ is the overlap of the flavor wave functions. The $I^{M_{L_A}m}_{M_{L_B}M_{L_C}}(\boldsymbol{p})$ stand for the spatial overlaps of  initial and final states, and relevant analytical expressions are listed in Appendix.
 Then, the decay width $\Gamma (A \rightarrow BC)$ can be  calculated directly
 \begin{eqnarray}
\Gamma= \pi^2\frac{p}{M^2_A}\frac{1}{2J_A+1}\sum_{M_{J_A},M_{J_B},M_{J_C}}|{\cal{M}}^{M_{J_A}M_{J_B}M_{J_C}}|^2,
\end{eqnarray}
where $p=|\boldsymbol{p}|=\frac{\sqrt{[M^2_A-(M_B+M_C)^2][M^2_A-(M_B-M_C)^2]}}{2M_A}$,
and $M_A$, $M_B$, and $M_C$ are the masses of the hadrons $A$, $B$, and $C$, respectively.

  In the  $^3P_0$ model, we need the quark pair creation strength $\gamma$, harmonic oscillator parameters in the orbital wave functions of initial and final states, and different types of quark masses to calculate the decays numerically. In the literature, the $\gamma = 0.4\times \sqrt{96 \pi}$ and $\alpha= 0.4~\rm{GeV}$ are widely used to investigate the strong decays of conventional mesons \cite{Godfrey:2015dva}, where the factor  $\sqrt{96 \pi}$ comes from different  conventions. For the conventional baryons, the $\alpha=0.4/\sqrt{2}$ GeV are commonly adopted, where the $\sqrt{2}$ originates from the different definitions of the relative coordinates \cite{Wang:2017hej,Lu:2018utx,Liang:2019aag,Lu:2019rtg,Zhong:2007gp}. In the compact tetraquarks, there is no experimental information to restrict these parameters via the known decay processes. In these $cq \bar c \bar q$ states, the $\rho_{1}$ and $\rho_{2}$ modes are similar to the excitations of conventional baryons, while the $\lambda$ mode is analogous to the conventional meson systems. Thus, we adopt  $\alpha_{\rho_1}$=$\alpha_{\rho_2}$=0.4/$\sqrt{2}$ GeV and $\alpha_\lambda=0.4$ GeV for the compact $cq \bar c \bar q$ states, $\alpha= 0.4~\rm{GeV}$ for light mesons, and $\gamma = 0.4\times \sqrt{96 \pi}$ for the overall strength. Also, the quark masses $m_{u/d}$=0.22 GeV, $m_{s}$=0.418 GeV, and $m_{c}$=1.628 GeV are employed in the calculations \cite{Godfrey:2015dva,Godfrey:1985xj,Capstick:1986ter}. Here, we expect to adopt as few parameters as possible to estimate the strong decays of  $cq \bar c \bar q$  tetraquarks semi-quantitatively and provide theoretical information for future explorations. 

 In present work, we plan to  study these  decay behaviors for the charmonium-like states $X(4700) $, $Z_c(4430)$,  $Y(4230)$, $Y(4360)$, $Y(4390)$, and $Y(4660)$. Combining experimental data and theoretical information, we  tentatively  regard  $X(4700) $ as a $cs \bar c \bar s(2S)$ state,  $Z_c(4430)$ as a  $cq \bar c \bar q(2S)$ state,  $Y(4230)$, $Y(4360)$, $Y(4390)$, $Y(4660)$ as the $cq \bar c \bar q(1P)$ states, and  $Y(4660)$ as the $cs \bar c \bar s(1P)$  states.  The possible quantum numbers of these states are listed in Table \ref{table1} . Here, for the  notation $\chi^{S_{12}S_{34}}_S $, the $S_{12}$, $S_{34}$, and $S$ are the  spin of $c_1 q_2$, $\bar c_3 \bar q_4$, and total spin, respectively. The explicit expressions of $\chi^{S_{12}S_{34}}_S $  are as follows
 \begin{eqnarray}
 	\chi^{00}_0={\left|{(c_1 q_2)}_0{(\bar c_3 \bar q_4)}_0\right\rangle} _0,
\end{eqnarray}
\begin{eqnarray}
 	\chi^{11}_0={\left|{(c_1 q_2)}_1{(\bar c_3 \bar q_4)}_1\right\rangle} _0,
\end{eqnarray}
\begin{eqnarray}
 	\chi^{01}_1={\left|{(c_1 q_2)}_0{(\bar c_3 \bar q_4)}_1\right\rangle} _1,
  \end{eqnarray}
\begin{eqnarray}
 	\chi^{10}_1={\left|{(c_1 q_2)}_1{(\bar c_3 \bar q_4)}_0\right\rangle} _1,
  \end{eqnarray}
\begin{eqnarray}
 	\chi^{11}_1={\left|{(c_1 q_2)}_1{(\bar c_3 \bar q_4)}_1\right\rangle} _1,
 	 \end{eqnarray}
\begin{eqnarray}
 	\chi^{11}_2={\left|{(c_1 q_2)}_1{(\bar c_3 \bar q_4)}_1\right\rangle} _2.
 \end{eqnarray}
Also, $Z_c(3985)$, $Z_c(4020)$, $Z_{cs}(3900)$, and $Z_{cs}(4000)$ are assumed to be the ground states of compact tetraquarks, which are all the $J^{P(C)}= 1^{+(-)}$ resonances. Under these assignments, the  $Z_c(3900)$ and $Z_{cs}(3985)$ resonances have the spin wave function of $(\chi^{10}_{1}-\chi^{01}_{1})/\sqrt{2}$, and the spin wave function for  $Z_c(4020)$ and $Z_{cs}(4000)$ resonances is $\chi^{11}_{1}$.
    
    \begin{table*}[htbp]
    \begin{center}
    	\caption{ \label{table1}Notations and quantum numbers of the initial  states. The superscripts and subscripts $S$, $A$, and $T$ in the states correspond to the the spins with 0, 1, and 2, respectively. }
    	\renewcommand{\arraystretch}{1.5}
    	\begin{tabular*}{18cm}{@{\extracolsep{\fill}}p{2.5cm}<{\centering}p{3.5cm}<{\centering}p{2.5cm}<{\centering}p{2.5cm}<{\centering}p{3cm}<{\centering}p{3cm}<{\centering}}
    		\hline\hline
    		State&$J^{PC}$ & $n_\lambda$ & $L_\lambda$&Spin &Candidate \\
            $X^{SS}_S$&$0^{++}$ &1&0&$\chi^{00}_{0}$&$\multirow{2}{*}{X(4700)}$\\
    		$X^{AA}_S$&$0^{++}$ &1&0&$\chi^{11}_{0}$&\\
    		\hline
    		$Y^{SS}_S$&$1^{--}$ &0&1&$\chi^{00}_{0}$&$\multirow{4}{*}{$\left(
    			\begin{array}{c}
    				Y(4660) \\ 
    				Y(4230) \\ 
    				Y(4360) \\ 
    				Y(4390)%
    			\end{array}%
    			\right)$}$ \\
    		$Y^{AA}_S$&$1^{--}$ &0&1&$\chi^{11}_{0}$& \\
    		$Y^{AA}_T$&$1^{--}$ &0&1&$\chi^{11}_{2}$&\\
    		$Y^{AS}_A$&$1^{--}$ &0&1&$\sqrt{\frac{1}{2}}(\chi^{10}_{1}+\chi^{01}_{1})$&\\
    		\hline
    		$Z^{AS}_A$&$1^{+-}$ &1&0&$(\chi^{10}_{1}-\chi^{01}_{1})/\sqrt{2}$&$\multirow{2}{*}{$Z_c(4430)$}$\\
    		$Z^{AA}_A$&$1^{+-}$ &1&0&$\chi^{11}_{1}$&\\

    		\hline \hline

    	\end{tabular*}
    \end{center}
\end{table*}

\section{STRONG DECAYS}{\label{low-lying}}.
\subsection{X(4700)}
\begin{table}[!htbp]
\begin{center}
\caption{ \label{asd1} Predictions of  the $ X(4700) \rightarrow  Z_{cs}(3985)/Z_{cs}(4000) K$ processes  in MeV.}
\renewcommand{\arraystretch}{1.5}
\begin{tabular*}{8.5cm}{@{\extracolsep{\fill}}p{4cm}<{\centering}p{2cm}<{\centering}p{2cm}<{\centering}}
\hline\hline
  \multirow{2}{*}{Decay mode} &  \multicolumn{2}{c}{X(4700)}  \\
  \cline{2-3}     
  &  $X^{SS}_S$ & $X^{AA}_S$  \\ 
 \hline
 $Z_{cs}(3985) K$ & 0.25 & 0.08 \\
$Z_{cs}(4000) K$  &0 &0.12\\
Total &0.25&0.20\\
\hline\hline
\end{tabular*}
\end{center}
\end{table}

In the literature, the $ X(4700)$ resonance was usually explained as a radially excited compact $cs \bar c \bar s$  tetraquark, $P-$wave conventional charmonium,  $D-$wave $cs \bar c \bar s$  tetraquark state, or a ground  tetraquark state
\cite{Chen:2010ze,Lu:2016cwr,Liu:2021xje,Deng:2017xlb,Ortega:2016hde,Chen:2016oma,Wang:2016gxp,Anwar:2018sol}. In the constituent quark model, there exist two $\lambda$-model $0^{++}$ $cs \bar c \bar s$ states, which are predicted to be close to each other. The partial decay widths of the $Z_{cs}(3985)  K$ and $Z_{cs}(4000) K$ channels are estimated and listed in  Table~\ref{asd1}. It can be seen that all of these partial widths are  relatively small, which suggests that the fall-apart process with two meson final states  may dominate. The small partial decay widths for $ X(4700)$ resonance may be caused by the radial $ 2 S$ excitations and the $P-$wave suppression of decay amplitude. Although the predicted partial decay widths are small, these decay channels also have the opportunities  to be observed in future experiments. Moreover, we find that the $Z_{cs}(4000) K$ decay mode is  forbidden at the  tree level diagrams under the $X^{SS}_S$ assignment, and this selection rule  is independent with  the parameters  of quark pair creation model.

\begin{figure}[!htbp]
\includegraphics[scale=0.8]{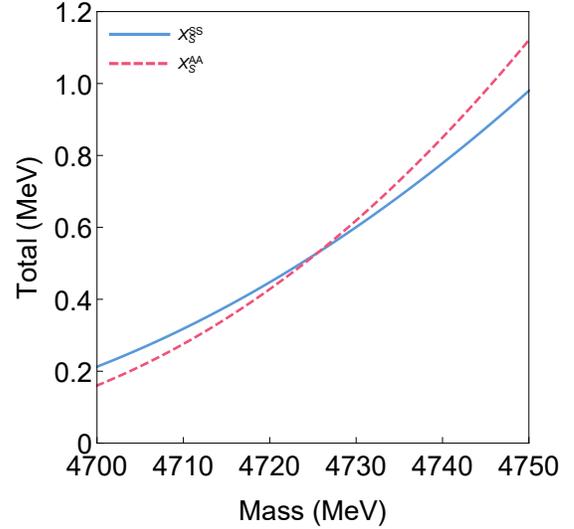}
\vspace{0.0cm} \caption{ The total widths of $Z_{cs}(3985)  K$ and $Z_{cs}(4000) K$  for the $X(4700)/X(4740)$ resonances  under different assignments versus the masses of initial states.}
\label{mass running}
\end{figure}

Recently, the LHCb Collaboration observed a new resonance $X(4740)$ in the $J/\psi \phi$ mass spectrum \cite{Ovsiannikova:2021zqp,LHCb:2020coc}, which is probably  the same state  as $X(4700)$. Also, this resonance  may be explained as another radially excited compact $0^{++}$ $cs \bar c \bar s$  tetraquark, or even a $2^{++}$ tetraquark. Here, we plot the  mass dependence of the partial decay widths for  two  $0^{++}$ states in Figure~\ref{mass running}. It can be seen that the total widths of  $Z_{cs}(3985) K$ and $Z_{cs}(4000) K$ channels  remain small when the mass varies from 4700 to 4750 MeV. We hope these estimations can provide valuable clue to better understand  the resonances $X(4700)$  and $X(4740)$, and  more  theoretical and experimental efforts are needed  to reveal  the relationship between   observed resonances and predicted states in quark model.

\subsection{$Z_c(4430)$}
\begin{table}[htb]\footnotesize
	\begin{center}
		\caption{ \label{asd2} Theoretical predictions of the strong decays for the $Z_c(4430)$ states in MeV.}
		\renewcommand{\arraystretch}{1.5}
		\begin{tabular*}{8.5cm}{@{\extracolsep{\fill}}p{4cm}<{\centering}p{2cm}<{\centering}p{2cm}<{\centering}}
			\hline\hline
 \multirow{2}{*}{Decay mode} &  \multicolumn{2}{c}{$Z_c(4430)$}  \\
  \cline{2-3}     
 &  $Z^{AS}_A$ & $Z^{AA}_A$    \\ 
 \hline
$Z_{c}(3900) \pi$ & $ 2.72\times 10^{-3}$ & $2.72\times 10^{-3}$ \\
$Z_{c}(4020) \pi$ & $8.60\times 10^{-2}$  & $8.60\times 10^{-2}$\\
Total         & $8.87\times 10^{-2}$  & $8.87\times 10^{-2}$\\
\hline\hline
\end{tabular*}
\end{center}
\end{table}
In the constituent quark model,  two $\lambda-$model $1^{+-}$   radially excited compact $cn \bar c \bar n$ tetraquarks  lie around 4.5 GeV. With these two assignments, the partial decay widths for the  $Z_c(4430)$ are calculated and presented in Table~\ref{asd2}. Similar to the $X(4700)$ case, the partial decay widths are  also small, owing to the radial excitations. Usually, the decay behaviors can help us to distinguish different assignments  for a new hadronic state. However, for the resonance $Z_c(4430)$ under two $1^{+-}$  assignments, its decays exhibit the same behaviors. This situation can be understood  by  the  Clebsch-Gordan coefficient in the  helicity amplitudes. Although these two assignments have different quantum numbers, it is found that the helicity amplitudes are same after tedious calculations. Because of the small partial decay widths and same decay behaviors, other decay channels may be more suitable to clarify the nature of  $Z_c(4430)$.

\subsection{Y(4230), Y(4360), and Y(4390)}
\begin{table}[htbp]
	\begin{center}
		\caption{ \label{asd3} Theoretical predictions of the strong decays for the Y(4230) states in MeV.}
		\renewcommand{\arraystretch}{1.5}
		\begin{tabular*}{8.5cm}{@{\extracolsep{\fill}}p{2.6cm}<{\centering}p{1.2cm}<{\centering}p{1.2cm}<{\centering}p{1.2cm}<{\centering}p{1.2cm}<{\centering}}
			\hline\hline
\multirow{2}{*}{Decay mode}  &  \multicolumn{4}{c}{$Y(4230)$}  \\
  \cline{2-5}    
			 & $Y^{SS}_S$ & $Y^{AA}_S$ & $Y^{AA}_T$ & $ Y^{AS}_A $  \\ 
			\hline
$Z_{c}(3900) \pi$ & 1.27  & 0.65 & 0.76  & 0.65\\
            $Z_{c}(4020) \pi$ &  0    & 0.58 & 0.96  & 0.58\\
            Total        & 1.27 & 1.23  & 1.72  &  1.23\\
			\hline\hline
		\end{tabular*}
	\end{center}
\end{table}

\begin{table}[htbp]
	\begin{center}
		\caption{ \label{asd4} Theoretical predictions of the strong decays for the Y(4360) states in MeV.}
		\renewcommand{\arraystretch}{1.5}
			\begin{tabular*}{8.5cm}{@{\extracolsep{\fill}}p{2.6cm}<{\centering}p{1.2cm}<{\centering}p{1.2cm}<{\centering}p{1.2cm}<{\centering}p{1.2cm}<{\centering}}
			\hline\hline
			
\multirow{2}{*}{Decay mode}  &  \multicolumn{4}{c}{$Y(4360)$}  \\
  \cline{2-5}    
			  & $Y^{SS}_S$ & $Y^{AA}_S$ & $Y^{AA}_T$ & $ Y^{AS}_A $  \\ 
			\hline
$Z_{c}(3900) \pi$ & 4.73  & 1.20 & 0.11  & 1.20\\
            $Z_{c}(4020) \pi$&  0    & 0.64 & 0.68  & 0.64\\
            Total        & 4.73 & 1.84  & 0.79 &  1.84\\
  
			\hline\hline
		\end{tabular*}
	\end{center}
\end{table}

\begin{table}[htbp]
	\begin{center}
		\caption{ \label{asd5} Theoretical predictions of the strong decays for the Y(4390) states in MeV.}
		\renewcommand{\arraystretch}{1.5}
			\begin{tabular*}{8.5cm}{@{\extracolsep{\fill}}p{2.6cm}<{\centering}p{1.2cm}<{\centering}p{1.2cm}<{\centering}p{1.2cm}<{\centering}p{1.2cm}<{\centering}}
			\hline\hline
			
\multirow{2}{*}{Decay mode}  &  \multicolumn{4}{c}{$Y(4390)$}  \\
  \cline{2-5}    
			  & $Y^{SS}_S$ & $Y^{AA}_S$ & $Y^{AA}_T$ & $ Y^{AS}_A $  \\ 
			\hline
$Z_{c}(3900) \pi$  & 5.64  & 1.47 & 0.22  & 1.47\\
            $Z_{c}(4020) \pi$&  0    & 0.63 & 0.50  & 0.63\\
            Total       & 5.64 & 2.10  & 0.72 &  2.10\\

			\hline\hline
		\end{tabular*}
	\end{center}
\end{table}

For the  $\lambda-$model $ P-$wave excitations, there are  four  $1^{--}$  $cn \bar c \bar n$ compact tetraquarks in the constituent quark model. Then we tentatively assign the $Y(4230)$, $Y(4360)$, and $Y(4390)$ as these  configurations, and calculated the partial decay widths of  $Z_{c}(3900) $ $\pi$ and $Z_{c}(4020) $  $\pi $ channels. The results are presented in \Cref{asd3,asd4,asd5}. The partial decay widths are predicted to be about $0.11 \sim 5.64$  MeV except the forbidden channels, which are significant enough to be   tested by forthcoming experiments. Moreover, the  $Y^{AA}_S$ and  $Y^{AS}_A$  assignments show the same decay behaviors for the pion emissions.

In the Ref.~\cite{Maiani:2021dzz}, the authors calculated  the process $Y(4230) \to Z_{c}(3900)  \pi$ in the tetraquark  scheme with the quark-pion axial vector interaction, and obtain a $ 8 ^{+10}_{-4}$ MeV partial decay width. When the same assignment is adopted, our present result is 0.76 MeV, which is  quite different from theirs. Under the  $ D\bar D_1 (2420) + H.c.$ . molecular scenario, we also investigated the decays $Y(4230) \to Z_{c}(3900) \pi$ in an effective Lagrangian approach~\cite{Chen:2016byt}, and obtain a  width of $3.15\pm0.45$ MeV. For the $Y(4390)$ resonance,   $ D^* \bar D_1  + H.c.$ molecular interpretation is also popular in the literature. In previous work \cite{Chen:2017abq}, the partial decay width of  $\pi^{+} \pi^{-} h_{c}$ process is estimated to be  0.74 to 0.85 MeV with an effective Lagrangian approach. By  subducting the branch ratio, one can get the width of $Y(4390) \to Z_{c}(4020) \pi $ process should be  around  ten  MeV, which is also larger than present results under compact tetraquark configurations. It should be emphasized that the $Y(4390) \to Z_c(3900)\pi$ is significant under compact tetraquark configurations, while the $Zc(3900)\pi$ mode should be suppressed for the $D^* \bar{D}_1 + H.c.$ molecular interpretation~\cite{Chen:2017abq}. Thus, the $Y(4390) \to Z_c(3900)\pi$ channel can help us to discriminate the nature of $Y(4390)$ and $Z_c(3900)$. More precisely  theoretical calculations and experimental information are needed to clarify their internal structures.
\begin{figure*}[!htbp]
\includegraphics[scale=0.7]{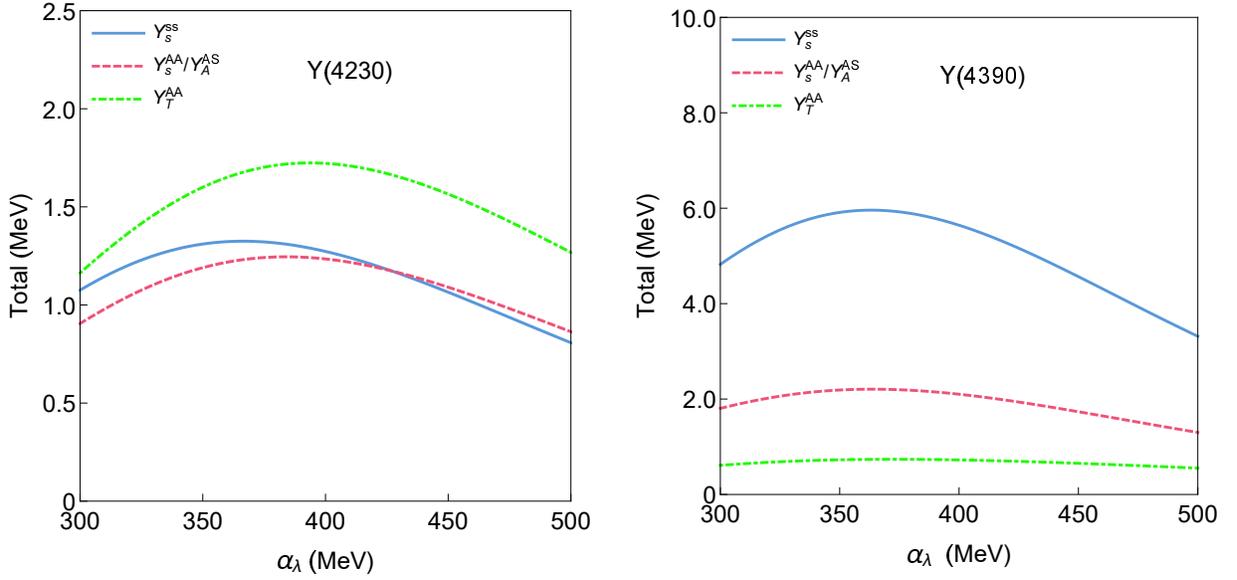}
\vspace{0.0cm} \caption{ The dependence on the harmonic oscillator parameter $\alpha_\lambda$ for $Y(4230)$ and $ Y(4390)$.}
\label{alpha}
\end{figure*}

\subsection{Y(4660)}
\begin{table}[!htbp]
	\begin{center}
		\caption{ \label{asd6} Theoretical predictions of the strong decays for the Y(4660) states in MeV.}
		\renewcommand{\arraystretch}{1.5}
			\begin{tabular*}{8.5cm}{@{\extracolsep{\fill}}p{2.6cm}<{\centering}p{1.2cm}<{\centering}p{1.2cm}<{\centering}p{1.2cm}<{\centering}p{1.2cm}<{\centering}}
			\hline\hline
			
\multirow{2}{*}{Decay mode}  &  \multicolumn{4}{c}{$Y(4660)$}  \\
  \cline{2-5}    
			& $Y^{SS}_S$ & $Y^{AA}_S$ & $Y^{AA}_T$ & $ Y^{AS}_A $ \\ 
\hline
			 $Z_{cs}(3985) K$ & 8.81  & 2.78 & 1.42  & 2.78 \\
            $Z_{cs}(4000) K$ &  0    & 2.89 & 2.46  & 2.89  \\
           Total        & 8.81 & 5.67  & 3.88 &  5.67   \\
        
		\hline\hline
		\end{tabular*}
	\end{center}
\end{table}
 
 The  $Y(4660)$ is a good candidate of the $P$-wave  $c  s \bar c \bar s$ compact tetraquarks.  This state was divided into two structures $Y(4660)$ and $Y(4630)$, however, some works indicated that $Y(4630)$ should be the same  structure as $Y(4660)$ \cite{Bugg:2008sk,Cotugno:2009ys,Guo:2010tk,Dai:2017fwx}. Here, we follow the assignment of  Review  of Particle Physics, which  suggests that only one  resonance $Y(4660)$ exits in this energy region~\cite{ParticleDataGroup:2022pth}.  The calculated  partial decay widths  are shown in Table \ref{asd6}. It is found that these  partial decay widths are significant, and the decay behaviors are similar to the non-strange  partners $Y(4230)$, $Y(4360)$, and $Y(4390)$.  Compared with the total width $62 ^{+9}_{-7}$ MeV, the  total branch ratios of  $ Z_{cs}(3985) K$ and $ Z_{cs}(4000) K$  channels are about $0.06 \sim 0.14$, which can be tested by  future experiments.

\subsection{Further discussions}

Our present  calculations are based on the resonances  $Z_c(3900)$/$Z_{cs}(3985)$  with  spin structure $(\chi^{10}_{1}-\chi^{01}_{1})/\sqrt{2}$, and   $Z_c(4020)$ and $Z_{cs}(4000)$ with  spin structure $\chi^{11}_{1}$. This is normal mass hierarchy for the $J^{P(C)}= 1^{+(-)}$ compact tetraquarks. However, these spin structures may be  
inversed in the realistic calculations, that is, $Z_c(3900)$/$Z_{cs}(3985)$  with  $\chi^{11}_{1}$, and   $Z_c(4020)$/$Z_{cs}(4000)$ with  $(\chi^{10}_{1}-\chi^{01}_{1})/\sqrt{2}$. Fortunately, because of the nearly equal masses of these states, the results  does not change significantly.  

In the constituent quark model, the commonly used interaction is the Cornell potential, where can reproduce the
low-lying spectra of heavy quarkonium well. With the realistic wave functions, one can obtain the effective  $\alpha$s that are predicted to be around 400 MeV. Considering the similarity between $\lambda$-mode compact tetraquarks and charmonium, we expect that the mass pattern and hierarchy can be reproduced. Indeed, some previous works \cite{Lu:2016cwr,Anwar:2018sol} based on the diquark-antiquark pictures show the masses and $\alpha$s in these ranges, and the experimental data of charmonium also supports the mass differences of compact tetraquarks. Moreover, we investigate the dependence on the harmonic oscillator parameter $\alpha_\lambda$ for $Y(4230)$ and $Y(4390)$ by fixing the relation $\alpha_\lambda = \sqrt{2} \alpha_\rho$ in Figure~\ref{alpha}. When the $\alpha_\lambda$ varies in a large reasonable range $300 \sim 500$ MeV, we can see that the decay widths have some changes but our conclusions remain.

Here, we apply the quark pair creation model to study the decay behaviors of compact tetraquarks, which is an  attempt to estimate the light meson emissions for the charmonium-like states. Also, investigations on these light meson emissions among compact tetraquarks are quite few both theoretically and  experimentally. In fact, this kind of processes is  important to study the relations of  compact tetraquarks and  needs more attention in future.

\section{SUMMARY}
In this work, we adopt the $^3P_0$ model to investigate light meson emissions of the  charmonium-like states $X(4700)$, $Z_c(4430)$, $Y(4230)$, $Y(4360)$, $Y(4390)$, and $Y(4660)$ under the  compact tetraquark  configurations.  It is found that the pion/kaon decay widths for the radially excited states are relatively small, while these partial decay widths for the $P$-wave excitations  are significant. Based on our calculations, we expect that the $Y(4390) \to Z_c(3900)/Z_c(4020)\pi$ decay channels are more likely to be measured by BESIII and Belle II Collaborations, which can provide valuable information to clarify the nature of $Y(4390)$, $Z_c(3900)$, and $Z_c(4020)$.

As the first strategy, our present  calculations are  preliminary. Here, we employ the widely accepted approach and relevant  parameters  to estimate the magnitudes of  transitions between these  charmonium-like states semi-quantitatively. Certainly, current  theoretical results and experimental data are not sufficient to  give  definitive  interpretations.  We hope that our  preliminary  explorations can provide information for  further researches, and more theoretical and  experimental  works are encouraged to discuss this topic.

\bigskip
\noindent
\begin{center}
	{\bf ACKNOWLEDGEMENTS}\\
\end{center}

We would like to thank Xian-Hui Zhong and Rui Chen for valuable discussions. This work is supported by the National Natural  Science Foundation of China under Grants No. 11705056, No. 12175037, No. 11947224, No. 11475192, No. 11975245, and No. U1832173. This work is also supported by the Key Project of Hunan Provincial Education Department under Grant No. 21A0039, the State Scholarship Fund of China Scholarship Council under Grant No. 202006725011, the Sino-German CRC 110 “Symmetries and the Emergence of Structure in QCD” project by NSFC
under the Grant No. 12070131001, the Key Research Program of Frontier Sciences, CAS, under the Grant No. Y7292610K1,
and the National Key Research and Development Program of China under Contracts No. 2020YFA0406300.

\begin{appendix}
\section*{Appendix: Spatial wave functions}
 \label{A}
\addcontentsline{toc}{chapter}{Appendix A Space wave function }

\setcounter{equation}{0}
\renewcommand\theequation{A.\arabic{equation}}

The harmonic oscillator wave functions for compact tetraquarks in momentum representation can be expressed as 
\begin{eqnarray}
	&&\psi _{n_{L}M_{L}}(n_{\rho _{1}},l_{\rho _{1}},m_{\rho _{1}},n_{\rho
		_{2}},l_{\rho _{2}},m_{\rho _{2}},n_{\lambda },l_{\lambda },m_{\lambda })=\nonumber\\
	&&(-1)^{n_{\rho _{1}}}(-i)^{l_{\rho _{1}}}p_{\rho _{1}}^{^{l_{\rho _{1}}}}\Bigg[%
	\frac{2n!}{\Gamma (n_{\rho _{1}}+l_{\rho _{1}}+\frac{3}{2})}\Bigg]^{\frac{1}{2}%
	}\times \Bigg(\frac{1}{\alpha _{\rho _{1}}}\Bigg)^{\frac{3}{2}+l_{\rho _{1}}} \nonumber\\
	&&(-1)^{n_{\rho _{2}}}(-i)^{l_{\rho _{2}}}p_{\rho _{2}}^{^{l_{\rho _{2}}}}\Bigg[%
	\frac{2n!}{\Gamma (n_{\rho _{2}}+l_{\rho _{2}}+\frac{3}{2})}\Bigg]^{\frac{1}{2}%
	}\times \Bigg(\frac{1}{\alpha _{\rho _{2}}}\Bigg)^{\frac{3}{2}+l_{\rho 2}} \nonumber\\
	&&(-1)^{n_{\rho _{\lambda }}}(-i)^{l_{\rho _{\lambda }}}p_{\rho _{\lambda
	}}^{^{l_{\rho _{\lambda }}}}\Bigg[\frac{2n!}{\Gamma (n_{\rho _{_{\lambda
		}}}+l_{\rho _{_{\lambda }}}+\frac{3}{2})}\Bigg]^{\frac{1}{2}}\times \Bigg(\frac{1}{%
		\alpha _{\rho _{_{\lambda }}}} \Bigg)^{\frac{3}{2}+l_{\rho _{\lambda }}} \nonumber\\
	&&Y_{l_{\rho _{1}}}^{m_{\rho _{1}}}(\vec p_{\rho _{1}})
	Y_{l_{\rho _{2}}}^{m_{\rho _{2}}}(\vec p_{\rho _{2}})Y_{l_{\rho
			_{\lambda }}}^{m_{\rho _{\lambda }}}(\vec p_{\lambda }%
	)L_{n_{\rho _{1}}}^{l_{\rho _{1}}+\frac{1}{2}}(p_{\rho _{1}}^{2}/\alpha
	_{\rho _{1}}^{2})
	L_{n_{\rho _{2}}}^{l_{\rho _{2}}+\frac{1}{2}}(p_{\rho
		_{2}}^{2}/ \alpha _{\rho _{2}}^{2})\nonumber \\
	&&L_{n_\lambda }^{l_\lambda+\frac{1}{2}} (p_\lambda^2 / \alpha _\lambda ^{2})
		\times \exp \Bigg(-\frac{%
		\vec p_{\rho _{1}}^{2}}{2\alpha _{\rho _{1}}^{2}}-\frac{%
		\vec p_{\rho _{2}}^{2}}{2\alpha _{\rho _{2}}^{2}}-\frac{%
		\vec p_{\rho _{\lambda}}^{2}}{2\alpha _{\rho _{\lambda}}^{2}}\Bigg),
\end{eqnarray}

\par{where} 

\begin{eqnarray}
\vec{p}_{\rho_1}=\frac{m_{2}\vec{p}_{1}- m_{1}\vec{p}_{2}}{m_{1}+ m_{2}},
\end{eqnarray}

\begin{eqnarray}
\vec{p}_{\rho_2}=\frac{m_{4}\vec{p}_{3}- m_{3}\vec{p}_{4}}{m_{3}+ m_{4}},
\end{eqnarray}

\begin{eqnarray}
\vec{p}_{\lambda}=\frac{(m_{3}+m_{4})(\vec{p}_{1}+\vec{p}_{2})- (m_{1}+m_{2})(\vec{p}_{3}+\vec{p}_{4})}{m_{1}+ m_{2}+ m_{3}+ m_{4}}.
\end{eqnarray}
Similarly, the harmonic oscillator wave function for ground mesons in momentum representation can be written as
\begin{eqnarray}
\psi \left(0,0,0\right)=\left( \frac{1}{\pi \alpha^{2}}\right) ^{\frac{3}{4}}\exp \left( -%
\frac{\vec{p}_{rel}^{2}}{2\alpha^{2}}\right),
\end{eqnarray}%
where $\vec{p}_{rel}$ represents the relative momentum between the quark and antiquark in the final mesons. 
\setlength{\parskip}{0.2cm plus4mm minus3mm}

In this paper, all the final states are ground states, that is, $n_{\rho_{B1}}=l_{\rho_{B1}}=n_{\rho_{B2}}=l_{\rho_{B2}}=n_{\lambda_ B}=l_{\lambda_ B}=n_{C}=L_{C}=0$. Also, we only consider the $\lambda$-mode excitations for initial states, and then $n_{\rho_{A1}}=l_{\rho_{A1}}=n_{\rho_{A2}}=l_{\rho_{A2}}=0$.  Here, we denote the spatial overlap integrals $I^{M_{L_A},m}_{M_{L_B},M_{L_C}}(\vec{p})$ as $\Pi( n_{\lambda _A}, l_{\lambda_A}, m_{\lambda_A}, m)$, and the relevant formulas for the low-lying states are present as follows. Define 
\begin{eqnarray}
c_{1} &=&\frac{m_{2}}{m_{1}+m_{2}}, 
\end{eqnarray}
\begin{eqnarray}
c_{2} &=&\frac{m_{1}+m_{2}}{2m_{1}+m_{2}+m_{5}}, 
\end{eqnarray}
\begin{eqnarray}
c_{3} &=&\frac{m_{1}}{m_{1}+m_{5}}, 
\end{eqnarray}
\begin{eqnarray}
c_{4} &=&\frac{m_{2}}{m_{2}+m_{5}}, 
\end{eqnarray}
\begin{eqnarray}
c_{5} &=&\frac{m_{1}}{m_{1}+m_{2}}-c_{3},
\end{eqnarray}
\begin{eqnarray}
\lambda _{1} &=&\frac{1}{\alpha _{\lambda }^{2}}+\frac{1}{2\alpha _{\rho
}^{2}}c_{5}^{2}+\frac{1}{2\alpha ^{2}}c_{1}^{2}, \\~~~~~~~~~~
\lambda _{2} &=&\frac{1}{\alpha _{\rho }^{2}}+\frac{1}{2\alpha ^{2}}, \\
\lambda _{3} &=&\frac{1}{2\alpha _{\rho }^{2}}c_{3}^{2}+\frac{1}{2\alpha
_{\lambda }^{2}}c_{2}^{2}+\frac{1}{2\alpha ^{2}}c_{4}^{2}, \\
\lambda _{4} &=&\frac{1}{\alpha _{\rho }^{2}}c_{3}c_{5}-\frac{1}{\alpha
_{\lambda }^{2}}c_{2}-\frac{1}{\alpha ^{2}}c_{1}c_{4}, \\
\lambda _{5} &=&\frac{1}{\alpha ^{2}}c_{1}-\frac{1}{\alpha _{\rho }^{2}}c_{5},
\\
\lambda _{6} &=&-\frac{1}{\alpha _{\rho }^{2}}c_{3}-\frac{1}{\alpha ^{2}}%
c_{4},
\end{eqnarray}
\begin{eqnarray}
f_{1} &=&\frac{\lambda _{5}}{2\lambda _{1}}, 
\end{eqnarray}
\begin{eqnarray}
f_{2} &=&\frac{\lambda _{4}}{2\lambda _{1}}, 
\end{eqnarray}
\begin{eqnarray}
f_{3} &=&\lambda _{2}-\frac{f_{1} \lambda _{5}}{2},
\end{eqnarray}
\begin{eqnarray}
f_{4} &=&\frac{\lambda _{6}-f_{2}\lambda _{5}}{2f_{3}}, 
\end{eqnarray}
\begin{eqnarray}
f_{5} &=&\lambda _{3}-\frac{f_{2} \lambda _{4}}{2}-f_{4}^{2}f_{3},
\end{eqnarray}
and then we can obtain the spatial overlaps integrals  straightforwardly
\begin{eqnarray}
\Pi (0,0,0,0)=\sqrt{\frac{3}{4\pi }}(1+f_{4}+c_{1}f_{2}-c_{1}f_{1}f_{4})
\left\vert \vec{P}\right\vert \Delta _{00},
\end{eqnarray}
\begin{eqnarray}
\Pi (1,0,0,0) &=&\Bigg[-2(1+f_{4}+c_{1}f_{2}-c_{1}f_{1}f_{4})(f_{2}-f_{1}f_{4})\left\vert \vec{P}\right\vert^2 \nonumber\\
&& -\frac{%
f_{1}(c_{1}f_{1}-1)}{f_{3}}-\frac{c_{1}}{\lambda _{1}}\Bigg]\Delta _{10}^{\lambda },
\end{eqnarray}
\begin{eqnarray}
\Pi (0,1,1,-1) =\Pi (0,1,-1,1)
=\Bigg[ \frac{f_{1}(c_{1}f_{1}-1)}{f_{3}}+\frac{c_{1}}{\lambda _{1}} \Bigg ]\Delta_{10}^{\lambda },  \nonumber\\
\end{eqnarray}
with
\begin{eqnarray}
\Delta _{00} &=& \Bigg( \frac{1}{\pi \alpha _{\rho }^{2}} \Bigg)^{3} \Bigg(\frac{1}{\pi \alpha
_{\lambda }^{2}}\Bigg)^{\frac{3}{2}} \Bigg(\frac{1}{\pi \alpha
^{2}}\Bigg)^{\frac{3}{4}} 
\Bigg(\frac{\alpha _{\rho }^{2}\pi ^{3}}{\lambda_{1}f_{3}} \Bigg)^{\frac{3}{2}} \nonumber\\
&&\times \exp \Bigg[-\Big(f_{5}\left\vert \vec{P}%
\right\vert ^{2}\Big) \Bigg], 
\end{eqnarray}
\begin{eqnarray}
\Delta _{10}^{\lambda } &=&-i\Bigg(\frac{1}{\pi \alpha _{\rho }^{2}}\Bigg)^{3} \Bigg(\frac{%
1}{\alpha _{\lambda }}\Bigg)^{\frac{5}{2}}\Bigg(\frac{\alpha _{\rho }^{2}\pi ^{3}}{%
\lambda _{1}f_{3}}\Bigg)^{\frac{3}{2}} \Bigg(\frac{1}{\pi \alpha _{\lambda }^{2}}\Bigg)^{%
\frac{3}{4}}\Bigg(\frac{1}{\pi \alpha
^{2}}\Bigg)^{\frac{3}{4}} \nonumber
\\
&&\times \exp \Bigg[-\Big(f_{5}\left\vert \vec{P}\right\vert ^{2}\Big) \Bigg]
\times \Bigg(\frac{8}{3\sqrt{\pi }}\Bigg)^{\frac{1}{2}}\Bigg(\frac{3}{8\pi }\Bigg).
\end{eqnarray}

\end{appendix}


\begin{thebibliography}{99}
	
	
	\bibitem{Hosaka:2016pey}
	A.~Hosaka, T.~Iijima, K.~Miyabayashi, Y.~Sakai and S.~Yasui,
	Exotic hadrons with heavy flavors: X, Y, Z, and related states,
	PTEP \textbf{2016}, 062C01(2016).
	
	
	\bibitem{Chen:2016qju}
	H.~X.~Chen, W.~Chen, X.~Liu and S.~L.~Zhu,
	The hidden-charm pentaquark and tetraquark states,
	Phys. Rept. \textbf{639}, 1-121 (2016).
	
	\bibitem{Lebed:2016hpi}
	R.~F.~Lebed, R.~E.~Mitchell and E.~S.~Swanson,
	Heavy-Quark QCD Exotica,
	Prog. Part. Nucl. Phys. \textbf{93}, 143-194 (2017).
	
	\bibitem{Dong:2017gaw}
	Y.~Dong, A.~Faessler and V.~E.~Lyubovitskij,
	Description of heavy exotic resonances as molecular states using phenomenological Lagrangians,
	Prog. Part. Nucl. Phys. \textbf{94}, 282-310 (2017).
	
	\bibitem{Guo:2017jvc}
	F.~K.~Guo, C.~Hanhart, U.~G.~Mei\ss{}ner, Q.~Wang, Q.~Zhao and B.~S.~Zou,
	Hadronic molecules,
	Rev. Mod. Phys. \textbf{90},015004 (2018).
	
	\bibitem{Olsen:2017bmm}
	S.~L.~Olsen, T.~Skwarnicki and D.~Zieminska,
	Nonstandard heavy mesons and baryons: Experimental evidence,
	Rev. Mod. Phys. \textbf{90},015003 (2018).
	
	\bibitem{Liu:2019zoy}
	Y.~R.~Liu, H.~X.~Chen, W.~Chen, X.~Liu and S.~L.~Zhu,
	Pentaquark and Tetraquark states,
	Prog. Part. Nucl. Phys. \textbf{107}, 237-320 (2019).
	
	\bibitem{Brambilla:2019esw}
	N.~Brambilla, S.~Eidelman, C.~Hanhart, A.~Nefediev, C.~P.~Shen, C.~E.~Thomas, A.~Vairo and C.~Z.~Yuan,
	The $XYZ$ states: experimental and theoretical status and perspectives,
	Phys. Rept. \textbf{873}, 1-154 (2020).
	
	
	\bibitem{Barabanov:2020jvn}
	M.~Y.~Barabanov, M.~A.~Bedolla, W.~K.~Brooks, G.~D.~Cates, C.~Chen, Y.~Chen, E.~Cisbani, M.~Ding, G.~Eichmann and R.~Ent, \textit{et al.}
	Diquark correlations in hadron physics: Origin, impact and evidence,
	Prog. Part. Nucl. Phys. \textbf{116}, 103835 (2021).
	
	
	\bibitem{Chen:2022asf}
	H.~X.~Chen, W.~Chen, X.~Liu, Y.~R.~Liu and S.~L.~Zhu,
	An updated review of the new hadron states,
	arXiv:2204.02649.
	
	\bibitem{Yue:2022gym}
	Z.~L.~Yue, M.~Y.~Duan, C.~H.~Liu, D.~Y.~Chen and Y.~B.~Dong,
	Hidden charm decays of X(4014) in a $D^*$ $\bar{D}^*$ molecule scenario,
	Phys. Rev. D \textbf{106},054008 (2022).
	
	
	\bibitem{Li:2013yla}
	G.~Li and X.~H.~Liu,
    Investigating possible decay modes of Y(4260) under $D_1(2420)\bar{D}$ + c.c. molecular state ansatz,
	Phys. Rev. D \textbf{88}, 094008 (2013).

	\bibitem{Ke:2013gia}
	H.~W.~Ke, Z.~T.~Wei and X.~Q.~Li,
	Is $Z_c(3900)$ a molecular state,
	Eur. Phys. J. C \textbf{73}, 2561 (2013).
	
	\bibitem{Chen:2019wjd}
	H.~X.~Chen,
   Decay properties of the $Z_c(3900)$ through the Fierz rearrangement,
	Chin. Phys. C \textbf{44}, 114003 (2020).
	
	\bibitem{Sundu:2018toi}
	H.~Sundu, S.~S.~Agaev and K.~Azizi,
	Resonance $Y(4660)$ as a vector tetraquark and its strong decay channels,
	Phys. Rev. D \textbf{98}, 054021(2018).


	\bibitem{Chen:2015fsa}
	W.~Chen, T.~G.~Steele, H.~X.~Chen and S.~L.~Zhu,
	$Z_c(4200)^+$ decay width as a charmonium-like tetraquark state,
	Eur. Phys. J. C \textbf{75}, 358(2015).
	
	
	\bibitem{Wang:2019iaa}
	Z.~G.~Wang
	Strong decays of the $Y(4660)$ as a vector tetraquark state in solid quark-hadron duality,
	Eur. Phys. J. C \textbf{79}, 184(2019).


	
	\bibitem{Liu:2014eka}
	X.~H.~Liu, L.~Ma, L.~P.~Sun, X.~Liu and S.~L.~Zhu,
 Resolving the puzzling decay patterns of charged $Z_c$ and $Z_b$ states,
	Phys. Rev. D \textbf{90}, no.7, 074020 (2014)
	doi:10.1103/PhysRevD.90.074020
	[arXiv:1407.3684 [hep-ph]].
	
	\bibitem{Wang:2018pwi}
	G.~J.~Wang, X.~H.~Liu, L.~Ma, X.~Liu, X.~L.~Chen, W.~Z.~Deng and S.~L.~Zhu,
   The strong decay patterns of $Z_c$ and $Z_b$ states in the relativized quark model,
	Eur. Phys. J. C \textbf{79}, 567 (2019).
	
	
	\bibitem{Xiao:2019spy}
	L.~Y.~Xiao, G.~J.~Wang and S.~L.~Zhu,
	Hidden-charm strong decays of the $Z_c$ states,
	Phys. Rev. D \textbf{101}, 054001(2020).
	
	
	
	\bibitem{Ferretti:2020civ}
	J.~Ferretti, E.~Santopinto, M.~N.~Anwar and Y.~Lu,
	Quark structure of the $\chi _{\mathrm{c}}(3P)$ and $X(4274)$ resonances and their strong and radiative decays,
	Eur. Phys. J. C \textbf{80}, 464(2020).
	
	
	\bibitem{Liu:2016nbm}
	X.~Liu, H.~W.~Ke, X.~Liu and X.~Q.~Li,
	Study of structures and dynamical decay mechanisms for multiquark systems,
	Phys. Rev. D \textbf{93}, 074013(2016).
	
	
	
	\bibitem{Wang:2022dfd}
	Z.~H.~Wang and G.~L.~Wang,
	Two-body strong decays of the 2P and 3P charmonium states,
	Phys. Rev. D \textbf{106},054037 (2022).
	
	
	\bibitem{Chen:2017abq}
	D.~Y.~Chen, C.~J.~Xiao and J.~He,
	Hidden-charm decays of Y(4390) in a hadronic molecular scenario,
	Phys. Rev. D \textbf{96}, 054017(2017).
	
	
	
	\bibitem{Chen:2016byt}
	D.~Y.~Chen, Y.~B.~Dong, M.~T.~Li and W.~L.~Wang,
	Pionic transition from Y(4260) to Z$_{c}$(3900) in a hadronic molecular scenario,
	Eur. Phys. J. A \textbf{52},310 (2016).
	
	
	
	
	
	
	

	
	
	\bibitem{LHCb:2016axx}
	R.~Aaij \textit{et al.} [LHCb],
	Observation of $J/\psi\phi$ structures consistent with exotic states from amplitude analysis of $B^+\to J/\psi \phi K^+$ decays,
	Phys. Rev. Lett. \textbf{118},022003 (2017).
	
	\bibitem{LHCb:2016nsl}
	R.~Aaij \textit{et al.} [LHCb],
	Amplitude analysis of $B^+\to J/\psi \phi K^+$ decays,
	Phys. Rev. D \textbf{95},012002 (2017).
	
	
	
	
	
	
	\bibitem{Chen:2010ze}
	W.~Chen and S.~L.~Zhu,
	The Vector and Axial-Vector Charmonium-like States,
	Phys. Rev. D \textbf{83}, 034010(2011).
	
	
	
	
	\bibitem{Lu:2016cwr}
	Q.~F.~L\"u and Y.~B.~Dong,
	X(4140) , X(4274) , X(4500) , and X(4700) in the relativized quark model,
	Phys. Rev. D \textbf{94}, 074007(2016).
	
	
	
	\bibitem{Liu:2021xje}
	X.~Liu, H.~Huang, J.~Ping, D.~Chen and X.~Zhu,
	The explanation of some exotic states in the $cs{\bar{c}}{\bar{s}}$ tetraquark system,
	Eur. Phys. J. C \textbf{81}, 950(2021).
	
	\bibitem{Deng:2017xlb}
	C.~Deng, J.~Ping, H.~Huang and F.~Wang,
	Hidden charmed states and multibody color flux-tube dynamics,
	Phys. Rev. D \textbf{98}, 014026(2018).
	
	
	
	
	
	\bibitem{Belle:2007hrb}
	S.~K.~Choi \textit{et al.} [Belle],
	Observation of a resonance-like structure in the $\pi^\pm \psi^\prime$ mass distribution in exclusive $B \to K \pi^{\pm} \psi^\prime$ decays,
	Phys. Rev. Lett. \textbf{100}, 142001 (2008).
	
	
	
	
	\bibitem{LHCb:2014zfx}
	R.~Aaij \textit{et al.} [LHCb],
	Observation of the resonant character of the $Z(4430)^-$ state,
	Phys. Rev. Lett. \textbf{112}, 222002 (2014).
	
	\bibitem{Belle:2014nuw}
	K.~Chilikin \textit{et al.} [Belle],
	Observation of a new charged charmoniumlike state in $\bar{B}^0 \rightarrow J/\psi K^- \pi^+$ decays,
	Phys. Rev. D \textbf{90}, 112009 (2014).
	
	
	\bibitem{ParticleDataGroup:2022pth}
	R.~L.~Workman \textit{et al.} [Particle Data Group],
	Review of Particle Physics,
	PTEP \textbf{2022}, 083C01(2022).
	
	\bibitem{BaBar:2005hhc}
	B.~Aubert \textit{et al.} [BaBar],
	Observation of a broad structure in the $\pi^+ \pi^- J/\psi$ mass spectrum around 4.26-GeV/c$^2$,
	Phys. Rev. Lett. \textbf{95}, 142001 (2005).
	
	
	\bibitem{Belle:2007dxy}
	C.~Z.~Yuan \textit{et al.} [Belle],
	Measurement of $e^+e^-\to \pi^+\pi^-J/\psi$cross-section via initial state radiation at Belle,
	Phys. Rev. Lett. \textbf{99}, 182004 (2007).
	
	
	
	
	
	
	
	
	
	
	\bibitem{BESIII:2016bnd}
	M.~Ablikim \textit{et al.} [BESIII],
	Precise measurement of the $e^+e^-\to \pi^+\pi^-J/\psi$ cross section at center-of-mass energies from 3.77 to 4.60 GeV,
	Phys. Rev. Lett. \textbf{118}, 092001 (2017).

	
	
	
	\bibitem{BESIII:2016adj}
	M.~Ablikim \textit{et al.} [BESIII],
	Evidence of Two Resonant Structures in $e^+ e^- \to \pi^+ \pi^- h_c$,
	Phys. Rev. Lett. \textbf{118},092002 (2017).
	
	\bibitem{BESIII:2020bgb}
	M.~Ablikim \textit{et al.} [BESIII],
	Observation of the $Y(4220)$ and $Y(4360)$ in the process $e^{+}e^{-} \to \eta J/\psi$,
	Phys. Rev. D \textbf{102}, 031101 (2020).
	
	
	\bibitem{BESIII:2022quc}
	M.~Ablikim \textit{et al.} [BESIII],
	Measurement of $e^{+}e^{-}\rightarrow\pi^{+}\pi^{-}D^{+}D^{-}$ cross sections at center-of-mass energies from 4.190 to 4.946 GeV,
	Phys. Rev. D \textbf{106}, 052012 (2022).
		
	\bibitem{BaBar:2006ait}
	B.~Aubert \textit{et al.} [BaBar],
	Evidence of a broad structure at an invariant mass of 4.32 $\rm{GeV/c^{2}}$ in the reaction $e^{+} e^{-} \to \pi^{+} \pi^{-} \psi_{2S}$ measured at BaBar,
	Phys. Rev. Lett. \textbf{98}, 212001 (2007).
	

	
	
	\bibitem{Belle:2007umv}
	X.~L.~Wang \textit{et al.} [Belle],
	Observation of Two Resonant Structures in $e^+ e^- \to \pi^+ \pi^- \psi(2S) $via Initial State Radiation at Belle,
	Phys. Rev. Lett. \textbf{99}, 142002 (2007).
	
	\bibitem{Belle:2014wyt}
	X.~L.~Wang \textit{et al.} [Belle],
	Measurement of $e^+e^- \to \pi^+\pi^-\psi(2S)$ via Initial State Radiation at Belle,
	Phys. Rev. D \textbf{91}, 112007 (2015).
	
	\bibitem{BaBar:2012hpr}
	J.~P.~Lees \textit{et al.} [BaBar],
	Study of the reaction $e^{+}e^{-}\to \psi(2S)\pi^{-}\pi^{-}$ via initial-state radiation at BaBar,
	Phys. Rev. D \textbf{89},111103 (2014).
	
	\bibitem{Belle:2008xmh}
	G.~Pakhlova \textit{et al.} [Belle],
	Observation of a near-threshold enhancement in the $e^+e^- \to \Lambda^+_c \Lambda^-_c$ cross section using initial-state radiation,
	Phys. Rev. Lett. \textbf{101}, 172001 (2008).
	
	
	
	
	
	
	
	
	
	
	
	
	
	
	\bibitem{Belle:2015hcs}
	Y.~L.~Han \textit{et al.} [Belle],
	Measurement of $e^+e^- \to \gamma\chi_{cJ}$ via initial state radiation at Belle,
	Phys. Rev. D \textbf{92}, 012011 (2015).
	
	
	
	
	
	
	
	
	
	\bibitem{Belle:2020wtd}
	S.~Jia \textit{et al.} [Belle],
	Evidence for a vector charmoniumlike state in $e^+e^- \to D^+_sD^*_{s2}(2573)^-+c.c.$,
	Phys. Rev. D \textbf{101}, 091101(2020).
	
	\bibitem{Belle:2019qoi}
	S.~Jia \textit{et al.} [Belle],
	Observation of a vector charmoniumlike state in $e^+e^- \to D^+_sD_{s1}(2536)^-+c.c.$,
	Phys. Rev. D \textbf{100}, 111103(2019).
	
	\bibitem{Micu:1968mk}
	L.~Micu,
	Decay rates of meson resonances in a quark model,
	Nucl. Phys. B \textbf{10}, 521-526 (1969).
	
	
	
	
	
	\bibitem{LeYaouanc:1988fx}
	A.~Le Yaouanc, L.~Oliver, O.~Pene and J.~C.~Raynal,
	Hardon Transitons in the quark model (Gordon and
	Breach, New York, 1988).
	
	\bibitem{LeYaouanc:1977gm}
	A.~Le Yaouanc, L.~Oliver, O.~Pene and J.~C.~Raynal,
	Why is $\psi(4.414) $ so narrow?,
	Phys. Lett. B \textbf{72}, 57-61 (1977).
	
	\bibitem{Roberts:1992esl}
	W.~Roberts and B.~Silvestre-Brac,
	General method of calculation of any hadronic decay in the $^3P_0$ model,
	Few Body Syst. \textbf{11}, 171-193 (1992).
	
	\bibitem{Ackleh:1996yt}
	E.~S.~Ackleh, T.~Barnes and E.~S.~Swanson,
	On the mechanism of open flavor strong decays,
	Phys. Rev. D \textbf{54}, 6811-6829 (1996).
	
	\bibitem{Barnes:1996ff}
	T.~Barnes, F.~E.~Close, P.~R.~Page and E.~S.~Swanson,
	Higher quarkonia,
	Phys. Rev. D \textbf{55}, 4157-4188 (1997).
	
	\bibitem{Barnes:2002mu}
	T.~Barnes, N.~Black and P.~R.~Page,
	Strong decays of strange quarkonia,
	Phys. Rev. D \textbf{68}, 054014 (2003).
	
	
	
	\bibitem{Chen:2016iyi}
	B.~Chen, K.~W.~Wei, X.~Liu and T.~Matsuki,
	Low-lying charmed and charmed-strange baryon states,
	Eur. Phys. J. C \textbf{77},  154 (2017).
	
	\bibitem{Lu:2016bbk}
	Q.~F.~L\"u, T.~T.~Pan, Y.~Y.~Wang, E.~Wang and D.~M.~Li,
	Excited bottom and bottom-strange mesons in the quark model,
	Phys. Rev. D \textbf{94},074012 (2016).
	
	\bibitem{Ferretti:2014xqa}
	J.~Ferretti, G.~Galat\`a and E.~Santopinto,
	Quark structure of the $X(3872)$ and $\chi_b(3P)$ resonances,
	Phys. Rev. D \textbf{90},054010 (2014).
	
	\bibitem{Godfrey:2015dva}
	S.~Godfrey and K.~Moats,
	Properties of Excited Charm and Charm-Strange Mesons,
	Phys. Rev. D \textbf{93},034035 (2016).
	
	\bibitem{Segovia:2012cd}
	J.~Segovia, D.~R.~Entem and F.~Fern\'andez,
	Scaling of the $^3P_0$ Strength in Heavy Meson Strong Decays,
	Phys. Lett. B \textbf{715}, 322-327 (2012).
 
	\bibitem{Zhao:2016qmh}
	Z.~Zhao, D.~D.~Ye and A.~Zhang,
	Nature of charmed strange baryons $\Xi_c(3055)$ and $\Xi_c(3080)$,
	Phys. Rev. D \textbf{94}, 114020 (2016).
	
	\bibitem{Chen:2007xf}
	C.~Chen, X.~L.~Chen, X.~Liu, W.~Z.~Deng and S.~L.~Zhu,
	Strong decays of charmed baryons,
	Phys. Rev. D \textbf{75}, 094017 (2007).
 
	\bibitem{Lu:2020ivo}
	Q.~F.~L\"u,
	Canonical interpretations of the newly observed $\Xi _c(2923)^0$, $\Xi _c(2939)^0$, and $\Xi _c(2965)^0$ resonances,
	Eur. Phys. J. C \textbf{80}, 921 (2020).
	
	\bibitem{Liang:2020hbo}
	W.~Liang and Q.~F.~L\"u,
	Strong decays of the newly observed narrow $\Omega _b$ structures,
	Eur. Phys. J. C \textbf{80}, 198 (2020).
	
	\bibitem{He:2021xrh}
	H.~Z.~He, W.~Liang, Q.~F.~L\"u and Y.~B.~Dong,
	Strong decays of the low-lying bottom strange baryons,
	Sci. China Phys. Mech. Astron. \textbf{64},261012 (2021).
	
\bibitem{Ader:1979bb}
J.~P.~Ader, B.~Bonnier and S.~Sood,
$B \bar{B}$ Widths of the Diquoniums Based on Quark Potential Model,
Z. Phys. C \textbf{5}, 85 (1980)


\bibitem{Roberts:1990ky}
W.~Roberts, B.~Silvestre- Brac and C.~Gignoux,
Baryon anti-baryon decays of four quark states,
Phys. Rev. D \textbf{41}, 182-194 (1990)

\bibitem{Liu:2016sip}
X.~Liu, H.~W.~Ke, X.~Liu and X.~Q.~Li,
Exploring open-charm decay mode $\Lambda _c\bar{\Lambda }_c$ of charmonium-like state $Y(4630)$,
Eur. Phys. J. C \textbf{76}, no.10, 549 (2016)
	
	
	\bibitem{Wang:2017hej}
	K.~L.~Wang, L.~Y.~Xiao, X.~H.~Zhong and Q.~Zhao,
	Understanding the newly observed $\Omega_c$ states through their decays,
	Phys. Rev. D \textbf{95}, 116010 (2017).
	
	
	\bibitem{Lu:2018utx}
	Q.~F.~L\"u, L.~Y.~Xiao, Z.~Y.~Wang and X.~H.~Zhong,
	Strong decay of $\Lambda _c(2940)$ as a $2P$ state in the $\Lambda _c$ family,
	Eur. Phys. J. C \textbf{78},599 (2018).
	
	\bibitem{Liang:2019aag}
	W.~Liang, Q.~F.~L\"u and X.~H.~Zhong,
	Canonical interpretation of the newly observed $\Lambda_b(6146)^0$ and $\Lambda_b(6152)^0$ via strong decay behaviors,
	Phys. Rev. D \textbf{100}, 054013 (2019).
	
	
	\bibitem{Lu:2019rtg}
	Q.~F.~L\"u and X.~H.~Zhong,
	Strong decays of the higher excited $\Lambda_Q$ and $\Sigma_Q$ baryons,
	Phys. Rev. D \textbf{101},014017 (2020).
	
	
	\bibitem{Zhong:2007gp}
	X.~H.~Zhong and Q.~Zhao,
	Charmed baryon strong decays in a chiral quark model,
	Phys. Rev. D \textbf{77}, 074008 (2008).
	
	
	
	
	
	
	
	
	
	
	\bibitem{Godfrey:1985xj}
	S.~Godfrey and N.~Isgur,
	Mesons in a Relativized Quark Model with Chromodynamics,
	Phys. Rev. D \textbf{32}, 189-231 (1985).
	
	\bibitem{Capstick:1986ter}
	S.~Capstick and N.~Isgur,
	Baryons in a relativized quark model with chromodynamics,
	Phys. Rev. D \textbf{34}, 2809-2835 (1986).
	
	
	
	
	\bibitem{Ortega:2016hde}
	P.~G.~Ortega, J.~Segovia, D.~R.~Entem and F.~Fern\'andez,
	Canonical description of the new LHCb resonances,
	Phys. Rev. D \textbf{94}, 114018 (2016).
	
	\bibitem{Chen:2016oma}
	H.~X.~Chen, E.~L.~Cui, W.~Chen, X.~Liu and S.~L.~Zhu,
	Understanding the internal structures of the $X(4140)$, $X(4274)$, $X(4500)$ and $X(4700)$,
	Eur. Phys. J. C \textbf{77},  160 (2017).
	
	\bibitem{Wang:2016gxp}
	Z.~G.~Wang,
	Scalar tetraquark state candidates: $X(3915)$, $X(4500)$ and $X(4700)$,
	Eur. Phys. J. C \textbf{77}, 78 (2017).


\bibitem{Anwar:2018sol}
M.~N.~Anwar, J.~Ferretti and E.~Santopinto,
Spectroscopy of the hidden-charm $[qc][\bar q \bar c]$ and $[sc][\bar s \bar c]$ tetraquarks in the relativized diquark model,
Phys. Rev. D \textbf{98},094015 (2018)
	
	
	
	\bibitem{Ovsiannikova:2021zqp}
	T.~Ovsiannikova [LHCb],
	Spectroscopy in Beauty Decays at the LHCb Experiment,
	Acta Phys. Polon. B \textbf{52},1047 (2021).
	
	\bibitem{LHCb:2020coc}
	R.~Aaij \textit{et al.} [LHCb],
	Study of $B^0_s \to J\psi \pi^+\pi^-K^+K^-$ decays,
	JHEP \textbf{02}, 024 (2021).
	
	
	\bibitem{Maiani:2021dzz}
	L.~Maiani, A.~D.~Polosa and V.~Riquer,
	Radiative and meson decays of $Y(4230)$ in flavor SU(3),
	Symmetry \textbf{13},751 (2021).
	
	
	
	\bibitem{Bugg:2008sk}
	D.~V.~Bugg,
	An Alternative fit to Belle mass spectra for $D \bar{D}, D^* \bar{D}^*$ and $\Lambda_c \bar{\Lambda}_c$,
	J. Phys. G \textbf{36}, 075002 (2009).
	
	\bibitem{Cotugno:2009ys}
	G.~Cotugno, R.~Faccini, A.~D.~Polosa and C.~Sabelli,
	Charmed Baryonium,
	Phys. Rev. Lett. \textbf{104}, 132005 (2010).
	
	\bibitem{Guo:2010tk}
	F.~K.~Guo, J.~Haidenbauer, C.~Hanhart and U.~G.~Meissner,
	Reconciling the X(4630) with the Y(4660),
	Phys. Rev. D \textbf{82}, 094008 (2010).
	
	
	
	\bibitem{Dai:2017fwx}
	L.~Y.~Dai, J.~Haidenbauer and U.~G.~Mei\ss{}ner,
	Re-examining the $X(4630)$ resonance in the reaction $e^+e^-\rightarrow \Lambda^+_c\bar\Lambda^-_c$,
	Phys. Rev. D \textbf{96}, 116001(2017).
	

	
	
\end{thebibliography}
\end{document}